\documentstyle[twoside,fleqn,espcrc2,curves]{article}
\input{epsf}

\newcommand{\AmS}{{\protect\the\textfont2
  A\kern-.1667em\lower.5ex\hbox{M}\kern-.125emS}}

\title{$f_B$ and the Heavy-light Spectrum from NRQCD}

\author{A. Ali Khan\address{Physics Department, The Ohio State  University,
Columbus, OH 43210, USA.}%
        \thanks{Talk presented by A. Ali Khan at LATTICE `97, Edinburgh, 
July 1997.}
}       
\begin{document}

\begin{abstract}
The present status of lattice calculations of the
$B$ spectrum and $f_B$, using NRQCD for the b quark, is discussed.
\end{abstract}

\maketitle

\section{INTRODUCTION}
$B$ physics is a subject of active research.
The theoretical and experimental understanding of the $B$ meson and b baryon 
spectrum is just beginning.
Weak matrix elements of $B$ mesons, e.g. $f_B$, $f_{B_s}$, $B_B$, and $B_{B_s}$
are being studied to determine fundamental parameters of the Standard 
Model. This review summarizes the progress made in calculating the $B$ 
spectrum and decay constants using the nonrelativistic QCD 
(NRQCD)~\cite{lepage} approach on the lattice. 

The advantage of NRQCD is that the rest mass term is removed from the
Lagrangian. Hence, large $O(M a)$ effects pose no problem, and one can simulate
b quarks directly on the lattice. Alternate approaches have
been reviewed by T.~Onogi~\cite{onogi} at this conference.

In heavy-light mesons,  NRQCD is equivalent to a $1/M$ expansion.
The rationale is that if the heavy quark 
is nonrelativistic ($p = Mv$), and the light
quark is relativistic with a momentum $p \sim \Lambda_{QCD}$, then 
momentum conservation in the meson rest frame gives
\begin{equation}
Mv \sim \Lambda_{QCD} .
\end{equation}
For B mesons, one has:
\begin{equation}
\Lambda_{QCD}/M \sim 0.1 .
\end{equation}
It is thus reasonable to include relativistic corrections in an 
expansion in powers of $\Lambda_{QCD}/M$. 
\begin{table}
\setlength{\tabcolsep}{0.1pc}
\begin{center}
\begin{tabular}{|l|l|l|l|l|}
\hline
\multicolumn{1}{|l|}{NRQCD} &
\multicolumn{1}{|l|}{group} &
\multicolumn{1}{|l|}{light quark} &
\multicolumn{1}{|c|}{$\beta$} &
\multicolumn{1}{|c|}{V} \\
\multicolumn{1}{|l|}{action} &
\multicolumn{1}{|l|}{} &
\multicolumn{1}{|l|}{action} &
\multicolumn{1}{|l|}{} &
\multicolumn{1}{|l|}{} \\
\hline
\multicolumn{5}{|c|}{quenched lattices} \\
\hline
$1/M$ & Hiroshima            & Wilson & 5.8 & $16^3\times 32$ \\
      & Draper {\em et al.}  & Wilson & 6.0 & $20^3\times 32$ \\
      & SGO                  & clover & 6.0 & $16^3\times 48$ \\
\hline
$1/M^2$ & Hiroshima          & Wilson & 5.8 & $16^3\times 32$ \\
        & GLOK               & clover & 5.7 & $12^3\times 24$ \\
        & GLOK               & clover & 6.0 & $16^3\times 48$ \\
\hline
\multicolumn{5}{|c|}{$n_f = 2$ staggered lattices from HEMCGC} \\
\hline
$1/M$   & SGO                & Wilson  & 5.6 & $16^3\times 48$ \\
        & SGO                & clover  & 5.6 & $16^3\times 48$ \\
\hline
\end{tabular}
\end{center}
\caption{NRQCD calculations of the $B$ spectrum and decay 
constants. For references see text.}
\label{tab:details}
\end{table}
The lowest terms in the expansion are as follows. 
The $1/M$ corrections to the static NRQCD 
Lagrangian density $L = D_t$ are 
\begin{equation}
H^{(1)} = - \frac{\vec{D}^2}{2M_0} - \frac{g}{2M_0}\vec{\sigma}\vec{B},
\end{equation}
and the $1/M^2$ corrections are 
\begin{eqnarray}
H^{(2)}\!\!& =& \!\frac{ig}{8M_0^2}\left(\vec{D}\vec{E} 
 - \vec{E}\vec{D}\right) \nonumber \\
\!  \!& - & \!\frac{g}{8M_0^2}\vec{\sigma}\left(\vec{D}\times\vec{E} - 
\vec{E}\times\vec{D}\right) - \frac{(\vec{D}^2)^2}{8M_0^3}.
\end{eqnarray}
In the following, actions which only include $H^{(1)}$ will be
referred to as $1/M$ actions, and those which include $H^{(2)}$, as
$1/M^2$ actions.  Note that $H^{(2)}$ also includes the first
relativistic correction to the kinetic energy which is formally
$O(1/M^3)$, but is expected to give a contribution of a similar size
as the spin-dependent interactions $O(1/M^2)$.  An overview of the
simulations covered is given in Table~\ref{tab:details}. For lack of
space I have not included a recent calculation on coarse lattices using
improved glue and light fermions~\cite{ma}.

In this review,  I will make comparisons to experiment and to
other lattice results where possible. To judge the reliability of the
predictions, I also evaluate systematic effects like the effect of
truncating the $1/M$ expansion, quenching errors, and the
dependence of the decay constant on lattice spacing.

\section{SPECTRUM}
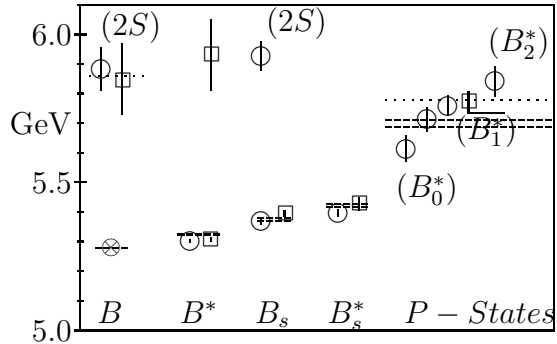
\begin{figure}[t]
\vspace{0.4cm}
\begin{center}
\setlength{\unitlength}{.0155in}
\begin{picture}(130,100)(20,500)

\put(15,500){\line(0,1){110}}
\put(175,500){\line(0,1){110}}
\multiput(13,500)(0,50){3}{\line(1,0){4}}
\multiput(14,500)(0,10){12}{\line(1,0){2}}
\put(12,500){\makebox(0,0)[r]{{\large5.0}}}
\put(12,550){\makebox(0,0)[r]{{\large5.5}}}
\put(12,600){\makebox(0,0)[r]{{\large 6.0}}}
\put(12,570){\makebox(0,0)[r]{{\large GeV}}}
\put(15,500){\line(1,0){160}}
\put(15,610){\line(1,0){160}}


     \put(25,510){\makebox(0,0)[t]{{\large $B$}}}
     \put(22,526){$\otimes$}
     \multiput(20,527.9)(3,0){4}{\line(1,0){2}}
     \put(22,588.3){\circle{6}}
     \put(22,588.3){\line(0,1){7.1}}
     \put(22,588.3){\line(0,-1){7.1}}
     \put(32,608){\makebox(0,0)[t]{{\large $(2S)$}}}
     \multiput(18,586)(3,0){7}{\line(1,0){0.5}}
     \put(29,581.5){$\!\!\Box$}
     \put(29,585){\line(0,1){12}}
     \put(29,585){\line(0,-1){12}}

     \put(55,510){\makebox(0,0)[t]{{\large $B^{*}$}}}
     \put(52,530.3){\circle{6}}
     \put(52,530.3){\line(0,1){0.5}}
     \put(52,530.3){\line(0,-1){0.5}}
     \multiput(48,532.6)(3,0){5}{\line(1,0){2}}
     \multiput(48,532.4)(3,0){5}{\line(1,0){2}}
\put(58.7,528.0){$\!\!\Box$}
\put(59,530.8){\line(0,1){.6}}
\put(59,530.8){\line(0,-1){.6}}
\put(59,590.3){$\!\!\Box$}
\put(59,593){\line(0,1){12}}
\put(59,593){\line(0,-1){12}}

     \put(80,510){\makebox(0,0)[t]{{\large $B_s$}}}
     \put(76,537){\circle{6}}
     \put(76,537){\line(0,1){1}}
     \put(76,537){\line(0,-1){1}}
     \multiput(75,538.1)(3,0){4}{\line(1,0){2}}
     \multiput(75,536.9)(3,0){4}{\line(1,0){2}}
     \put(76,592.7){\circle{6}}
     \put(76,592.7){\line(0,1){4.8}}
     \put(76,592.7){\line(0,-1){4.8}}
     \put(88,609){\makebox(0,0)[t]{{\large $(2S)$}}}
     \put(84,536.6){$\!\!\Box$}
     \put(84,539.4){\line(0,1){1}}
     \put(84,539.4){\line(0,-1){1}}
     \multiput(75,538.1)(3,0){4}{\line(1,0){2}}
     \multiput(75,536.9)(3,0){4}{\line(1,0){2}}

     \put(105,510){\makebox(0,0)[t]{{\large $B^{*}_s$}}}
     \put(102,539.8){\circle{6}}
     \put(102,539.8){\line(0,1){1.1}}
     \put(102,539.8){\line(0,-1){1.1}}
     \multiput(98,542.8)(3,0){5}{\line(1,0){2}}
     \multiput(98,541.6)(3,0){5}{\line(1,0){2}}
     \put(109,540.0){$\!\!\Box$}
     \put(109,542.7){\line(0,1){2.0}}
     \put(109,542.7){\line(0,-1){2.0}}

     \put(150,510){\makebox(0,0)[t]{{\large $P-States$}}}
     \put(155,584.2){\circle{6}}
     \put(155,584.2){\line(0,1){5}}
     \put(155,584.2){\line(0,-1){5}}
     \put(163,597){\makebox(0,0){{\large $(B^*_2)$}}}
     \put(125,561.5){\circle{6}}
     \put(125,561.5){\line(0,1){4.5}}
     \put(125,561.5){\line(0,-1){4.5}}
     \put(132,553.0){\makebox(0,0)[t]{{\large $(B^*_0)$}}}
     \put(132,571.3){\circle{6}}
     \put(132,571.3){\line(0,1){4}}
     \put(132,571.3){\line(0,-1){4}}
     \put(139,575.9){\circle{6}}
     \put(139,575.9){\line(0,1){3.4}}
     \put(139,575.9){\line(0,-1){3.4}}
     \put(146,574.55){$\!\!\Box$}
     \put(146,577.5){\line(0,1){3.4}}
     \put(146,577.5){\line(0,-1){3.4}}
     \put(152,568){\makebox(0,0){{\large $(\overline{B^*_1})$}}}
     \multiput(118,568.6)(3,0){19}{\line(1,0){2}}
     \multiput(118,571.0)(3,0){19}{\line(1,0){2}}
     \multiput(118,577.9)(3,0){19}{\line(1,0){0.5}}

\end{picture}
\end{center}
\vspace{-0.4cm}
\caption{The $B$ meson spectrum from lattice NRQCD. Circles denote quenched
results, squares, results from $n_f = 2$ lattices.
Dashed lines denote experimental error bounds as given 
by the Particle Data Book, dotted lines the central values of first
experimental measurements of these states. }
\vspace{-0.3cm}
\label{fig:messpec}
\end{figure}

Using nonrelativistic b quarks and clover light quarks on the lattice, it is
possible to reproduce the presently known general features of the $B$ spectrum.
An overview of the the meson spectrum at a lattice spacing 
$a^{-1} \sim 2$ GeV is presented in Fig.~\ref{fig:messpec}. The plot shows
results from the most comprehensive calculations on quenched ($1/M^2$ action,
GLOK~\cite{stlouis}) and dynamical ($1/M$ action, SGO~\cite{saraprogress}) 
configurations. 
The $B_s - B_d$
splitting agrees with experiment, also the $B(2S)$ agrees well with the first
experimental candidate~\cite{DELPHI}. 

For the $P$ wave states, two kinds of 
experimental signals have been found. A $B^{(\ast)}
\pi$ resonance has been established which is expected to be a superposition 
of various $P$ states.
The states with light quark angular momentum $j_l = 1/2$ ($B^\ast_0$ and
$B^\ast_1$) and $j_l = 3/2$ ($B^{\ast\prime}_1$ and $B^\ast_2$) are 
expected to form doublets. The splittings within these doublets are given by 
the coupling of the heavy quark spin to $j_l$.  The second experimental
signal is a narrow $B\pi\pi$ resonance which has probably 
$j_l = 3/2$~\cite{DELPHI}.

Lattice NRQCD predicts a $B^\ast_2 - B^\ast_0$ splitting of the order
of 200 MeV (an effect of $\sim 4 \sigma$). The $B^{\ast}_1$ and 
$B^{\ast\prime}_1$
cannot yet be resolved since each of the two operators used in the simulation 
for the $j = 1$ states project only on a superposition of these states, 
called  $\overline{B^{\ast}_1}$ in Fig.~\ref{fig:messpec}. For $n_f = 2$
lattices, there
is presently only a $P$ state signal  in the channel that (similar to the 
quenched data point on the left of it) corresponds to the
$^1P_1$ state in heavy-heavy mesons. Given the 
limitations that
apply to both the experimental and lattice results on $P$
states, my conclusion is that there is qualitative agreement. 

A study of the effect of the $1/M^2$ 
terms in the action on some spectral quantities is  
ongoing~\cite{heinfuture}. A low statistics comparison between $1/M$ and $1/M^2$ actions
on quenched configurations at $\beta = 6.0$~\cite{saraprogress,stlouis}, suggests that 
there is no significant difference.  Assuming that this is true, a comparison of the 
spectrum on quenched and the $n_f = 2$ lattices from HEMCGC, as 
shown in Fig.~\ref{fig:messpec}, further suggests that quenching errors are small too.

Lastly, it is important to study the dependence of the results on
the lattice spacing $a$.  Preliminary results of a comparison 
between quenched lattices at $\beta = 5.7$ and $6.0$ show no
significant scaling violations~\cite{hein}. 

\subsection{The Hyperfine Splitting}
\begin{figure}[htb]
\vspace{-5.94cm}
\centerline{\hspace{0.5cm}\epsfxsize=10cm
\epsfbox{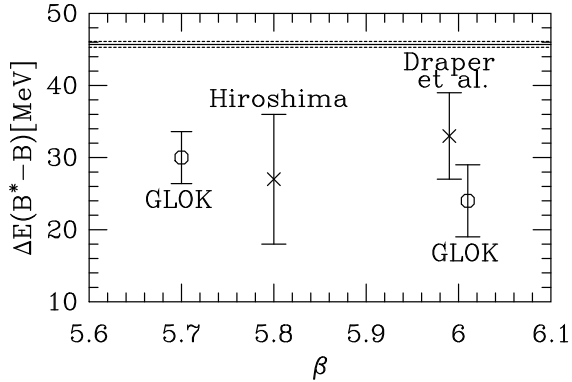}}
\vspace{-2.4cm}
\caption{NRQCD results for the $B$ hyperfine splitting. Crosses use Wilson 
light quarks, circles tadpole-improved clover light quarks. The solid and
dotted lines denote the experimental value with its error bounds.}
\label{fig:hypscal}
\vspace{-0.65cm}
\end{figure}

The $B^\ast-B$ splitting is of particular interest since it is expected to 
be sensitive 
to various effects such as discretization errors, quenching and the tuning of
the parameters in the heavy quark action. The results shown in 
Fig.~\ref{fig:hypscal} are $\sim 20$ MeV too low. 
The question arises whether the results improve for smaller $a$. 
Fig.~\ref{fig:hypscal} shows results using Wilson and clover light
quarks at various lattice spacings, with $a$ determined from $m_\rho$.
With present errors, I conclude that there is reasonable scaling for
clover light quark data at $\beta = 5.7$ and $6.0$, as well as for
Wilson light quark data at $\beta = 5.8$ and $6.0$. Also, there is no
obvious difference between clover and Wilson results.

If the $B$ hyperfine splitting depends on the heavy quark wavefunction
at the origin, one would expect quenching effects to be
significant. Comparing results on quenched and dynamical ($n_f = 2$)
configurations (Refs.~\cite{saraprogress,stlouis}, does not show any
significant effect.  One possibility is that the physical picture of a
nonrelativistic wavefunction is incorrect in $B$ mesons.  However, the
sea quark mass in the HEMCGC configurations might be too large, so
modern dynamical simulations with high statistics are 
needed to settle
this issue.  Another quantity whose effect on the hyperfine splitting
needs to be studied are the coefficients of the spin-dependent terms
in the action. A perturbative calculation by
H. Trottier~\cite{trottier} is under way.
\subsection{Comparison to other lattice calculations}
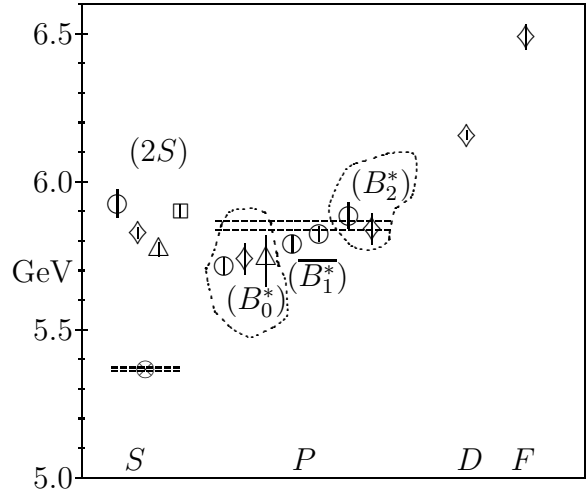
\begin{figure}[t]
\vspace{2.36cm}
\begin{center}
\setlength{\unitlength}{.0155in}
\begin{picture}(160,100)(10,500)
\put(15,500){\line(0,1){160}}
\put(185,500){\line(0,1){160}}
\multiput(13,500)(0,50){4}{\line(1,0){4}}
\multiput(14,500)(0,10){17}{\line(1,0){2}}
\put(12,500){\makebox(0,0)[r]{{\large5.0}}}
\put(12,550){\makebox(0,0)[r]{{\large5.5}}}
\put(12,600){\makebox(0,0)[r]{{\large 6.0}}}
\put(12,650){\makebox(0,0)[r]{{\large 6.5}}}
\put(12,570){\makebox(0,0)[r]{{\large GeV}}}
\put(15,500){\line(1,0){170}}
\put(15,660){\line(1,0){170}}


     \put(33,510){\makebox(0,0)[t]{{\large $S$}}}
     \multiput(25,536.2)(3,0){8}{\line(1,0){2}}
     \multiput(25,537.2)(3,0){8}{\line(1,0){2}}
     \put(33,534.4){$\otimes$}
     \put(27,592.6){\circle{6}}
     \put(27,592.6){\line(0,1){4.7}}
     \put(27,592.6){\line(0,-1){4.7}}
     \put(30.4,580.3){$\diamondsuit$}
     \put(34,582.8){\line(0,1){1.8}}
     \put(34,582.8){\line(0,-1){1.8}}
     \put(37,574.9){$\triangle$}
     \put(41,577.1){\line(0,1){2.3}}
     \put(41,577.1){\line(0,-1){2.3}}
      \put(48,587.3){$\!\!\Box$}
     \put(48,590.0){\line(0,1){2.3}}
     \put(48,590.0){\line(0,-1){1.9}}
     \put(41,615){\makebox(0,0)[t]{{\large $(2S)$}}}

     \put(90,510){\makebox(0,0)[t]{{\large $P$}}}
     \put(105,588.5){\circle{6}}
     \put(105,588.5){\line(0,1){4.4}}
     \put(105,588.5){\line(0,-1){4.4}}
     \put(109.4,581.6){$\diamondsuit$}
     \put(113,584.1){\line(0,1){5.1}}
     \put(113,584.1){\line(0,-1){5.1}}
	\curvedashes[0.2mm]{0,1,3}
	\closecurve(98.6,592,99,587.5,101,584,106.5,578,112,577,116,578,
         119,581,120,588,121,591,124,593,126.5,
         595,128,603,128,607,127,609.5,124,610,118,609.5,112,607.5,
         107,606,104,604)
     \put(116,599){\makebox(0,0){{\large $(B^*_2)$}}}
     \put(63,571.4){\circle{6}}
     \put(63,571.4){\line(0,1){2.9}}
     \put(63,571.4){\line(0,-1){2.9}}
     \put(66.4,571.4){$\diamondsuit$}
     \put(70,573.9){\line(0,1){5.1}}
     \put(70,573.9){\line(0,-1){5.1}}
     \put(73.2,571.8){$\triangle$}
     \put(77,573.2){\line(0,1){8.6}}
     \put(77,573.2){\line(0,-1){8.6}}
	\curvedashes[0.2mm]{0,1,3}
	\closecurve(57,571,57.6,566,59,560,60,556,62,554,65,551,
        68,549,70,548,72,547.5,74,548,76,549,80,551,84,554,84.5,556,
         84,565,83.5,568,82.5,570,81,573,81,576,80,580,79,586,
         78,587.5,77,589.5,
         76,590,73,591,65,590,63,587,62,585,61,581,59.5,576)
     \put(74,565){\makebox(0,0)[t]{{\large $(B^*_0)$}}}
     \put(94.5,568){\makebox(0,0){{\large $(\overline{B^*_1})$}}}
     \put(86,579.0){\circle{6}}
     \put(86,579.0){\line(0,1){2.9}}
     \put(86,579.0){\line(0,-1){2.9}}
     \put(95,582.2){\circle{6}}
     \put(95,582.2){\line(0,1){2.5}}
     \put(95,582.2){\line(0,-1){2.5}}
     \multiput(60,583.8)(3,0){20}{\line(1,0){2}}
     \multiput(60,586.8)(3,0){20}{\line(1,0){2}}

     \put(146,510){\makebox(0,0)[t]{{\large $D$}}}
     \put(141.4,613.3){$\diamondsuit$}
     \put(145,615.8){\line(0,1){1.4}}
     \put(145,615.8){\line(0,-1){1.4}}

     \put(164,510){\makebox(0,0)[t]{{\large $F$}}}
     \put(161.4,646.5){$\diamondsuit$}
     \put(165,649.0){\line(0,1){4.2}}
     \put(165,649.0){\line(0,-1){4.2}}

\end{picture}
\vspace{-0.3cm}
\end{center}
\caption{Compilation of lattice results for excited $B_s$ mesons. Circles 
are  NRQCD results, diamonds~\protect\cite{peisa} and 
triangles~\protect\cite{duncan} show static heavy quarks. The square shows
tadpole-improved clover heavy quarks~\protect\cite{jim}. Results for
the $B_0^\ast$ and the $B_2^\ast$ are grouped together wih dotted circles. The 
meaning  of the dashed lines is the same as in Fig.~\protect\ref{fig:messpec}. 
At the bottom of the figure, the orbital quantum numbers of the respective
states are shown.}
\label{fig:excited_Bs}
\vspace{-0.2cm}
\end{figure}

Fig.~\ref{fig:excited_Bs} shows data from NRQCD and other lattice
methods along with a preliminary experimental result for a $P$ state
excited $B_s$ meson.  In some of the calculations only an approximate
value for the strange quark mass is used~\cite{peisa,duncan}. Also,
the calculations were done at slightly different $\beta$ values
between 5.9 and 6.1. (The lattice spacing was set using either 
$m_\rho$ with clover quarks or $\sqrt{\sigma}$. These give 
roughly consistent values for a fixed $\beta$ in this $\beta$ range.)
Because of these limitations, a check of the expected $1/M$ corrections ($O(50)$ MeV) to the static 
limit cannot be made with these data. 
Lastly, I would like to mention that first results for $D$ and $F$
wave states are reported in Ref.~\cite{peisa}.
\subsection{b Baryons}
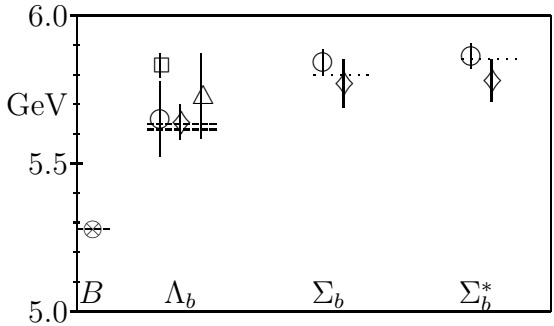
\begin{figure}[tb]
\begin{center}
\hspace{0.5cm}
\setlength{\unitlength}{.0155in}
\begin{picture}(130,100)(25,500)
\put(15,500){\line(0,1){100}}
\put(175,500){\line(0,1){100}}
\multiput(13,500)(0,50){3}{\line(1,0){4}}
\multiput(14,500)(0,10){11}{\line(1,0){2}}
\put(12,500){\makebox(0,0)[r]{{\large5.0}}}
\put(12,550){\makebox(0,0)[r]{{\large5.5}}}
\put(12,600){\makebox(0,0)[r]{{\large 6.0}}}
\put(12,570){\makebox(0,0)[r]{{\large GeV}}}
\put(15,500){\line(1,0){160}}
\put(15,600){\line(1,0){160}}



     \put(20,510){\makebox(0,0)[t]{{\large $B$}}}
     \put(20,525.4){$\!\!\otimes$}
     \multiput(15,527.9)(3,0){4}{\line(1,0){2}}

     \put(50,510){\makebox(0,0)[t]{{\large $\Lambda_b$}}}
     \put(43,565.1){\circle{6}}
     \put(43,565.1){\line(0,1){12.6}}
     \put(43,565.1){\line(0,-1){12.6}}
     \multiput(39,561.5)(3,0){8}{\line(1,0){2}}
     \multiput(39,563.3)(3,0){8}{\line(1,0){2}}
     \put(46.9,561.4){$\diamondsuit$}
     \put(50,564){\line(0,1){6}}
     \put(50,564){\line(0,-1){6}}
     \put(53.8,570.1){$\triangle$}
     \put(57,572.8){\line(0,1){14.5}}
     \put(57,572.8){\line(0,-1){14.5}}
     \put(43.3,580.2){$\!\!\Box$}
     \put(43,583){\line(0,1){4}}
     \put(43,583){\line(0,-1){4}}

     \put(100,510){\makebox(0,0)[t]{{\large $\Sigma_b$}}}
     \put(98,584.3){\circle{6}}
     \put(98,584.3){\line(0,1){4.3}}
     \put(98,584.3){\line(0,-1){4.3}}
     \multiput(95,579.7)(3,0){7}{\line(1,0){0.5}}
     \put(104.9,574.5){$\!\!\diamondsuit$}
     \put(105,577){\line(0,1){8}}
     \put(105,577){\line(0,-1){8}}

     \put(150,510){\makebox(0,0)[t]{{\large $\Sigma_b^*$}}}
     \put(148,586.2){\circle{6}}
     \put(148,586.2){\line(0,1){4.3}}
     \put(148,586.2){\line(0,-1){4.3}}
     \multiput(145,585.3)(3,0){7}{\line(1,0){0.5}}
     \put(154.9,575.5){$\!\!\diamondsuit$}
     \put(155,578){\line(0,1){7}}
     \put(155,578){\line(0,-1){7}}

\end{picture}
\end{center}
\vspace{-0.3cm}
\caption{Lattice results on baryons containing one b and two light quarks.
Only lattice calculations with moving (non-static) b quarks have been 
considered.
Circles denote NRQCD b quarks~\protect\cite{stlouis}, diamonds show clover
($c_{SW} = 1$) heavy quarks~\protect\cite{UKQCD_baryons}, and the triangle,  
Wilson heavy quarks~\protect\cite{wupp_baryons}. The square is from a SGO NRQCD
calculation with $n_f = 2$~\protect\cite{saraprogress}. The 
meaning  of the lines is the same as in Fig.~\protect\ref{fig:messpec}.}
\label{fig:bar}
\end{figure}

Lattice results for b baryons with non-static b quarks are summarized in 
Fig.~\ref{fig:bar}. For the $\Lambda_b$ the quenched lattice calculations 
using NRQCD~\cite{stlouis}, tree-level clover~\cite{UKQCD_baryons}, and plain 
Wilson~\cite{wupp_baryons} heavy quarks agree within errors with experiment 
and with each other. The $\Lambda_b$ from Ref.~\cite{saraprogress} is 
significantly heavier than the experimental 
value and heavier than the quenched NRQCD result. This  needs further study.

For the $\Sigma_b$ and $\Sigma^\ast_b$ baryons there are
only preliminary experimental results available~\cite{DELPHIbar}, but they
are in agreement with the available lattice results (see Fig.~\ref{fig:bar}). 
Ref.~\cite{UKQCD_baryons} also has results for strange b baryons. 
\section{\label{sec:decay} DECAY CONSTANTS}
At tree level, the nonrelativistic current, correct through $O(1/M^2)$, is 
given by:
\begin{eqnarray}
A_0 \!&=& \overline{q}\gamma_0\gamma_5 Q \label{eq:m0} \\
 \!&-& \frac{1}{M_0}\overline{q}\gamma_0\gamma_5\left(\vec{\gamma}
\vec{D}\right) Q \label{eq:m1} \\
\!&+&\frac{1}{8M_0^2}\bar{q}\gamma_0\gamma_5\left(\vec{D}^2 + 
g\vec{\Sigma} \vec{B} - 2i\vec{\alpha}\vec{E}\right)Q \label{eq:m2},
\end{eqnarray}
where $\vec{\alpha} = \gamma_0\vec{\gamma}$. 

The question is whether the 
truncation of the $1/M$ expansion at this order is sensible. At present, 
the effects of including both $1/M$ and $1/M^2$ 
corrections have been studied in two calculations: Ref.~\cite{hiroshima}
with Wilson light quarks and Ref.~\cite{stlouis} with clover
light quarks. First I examine the effect of the $1/M^2$ terms in the 
action on the matrix elements of the zeroth order 
current (Eq.~(\ref{eq:m0})): Comparing the $1/M$ and $1/M^2$ actions with
clover light quarks at $\beta = 6.0$, one finds a difference of 
$\sim 7\%$~\cite{stlouis,sgo_quenched}. 
However, this is only a $\sim 2\sigma$ effect as the results 
from~\cite{sgo_quenched} are low statistics. 
Similarly, the Hiroshima group
finds the effect of including the $1/M^2$ corrections 
to be $< 0.5\%$~\cite{hiroshima}. 

\begin{figure}[t]
\vspace{-5.94cm}
\centerline{\hspace{0.5cm}\epsfxsize=10cm
\epsfbox{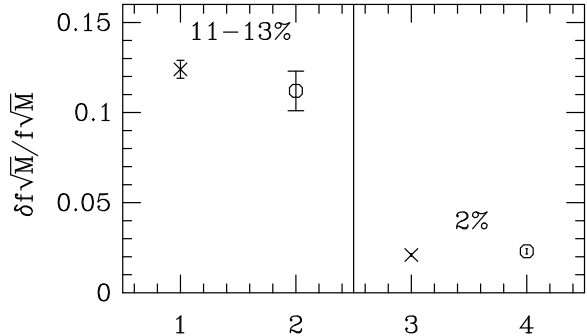}}
\vspace{-2.5cm}
\caption{Ratio of the unrenormalized current corrections (left, $O(1/M)$,
right, $O(1/M^2)$) to the uncorrected 
current. Crosses use Wilson light quarks at 
$\beta = 5.8$~\protect\cite{hiroshima}, circles, clover light quarks at
$\beta = 6.0$ (GLOK). For the x axis, an arbitrary numbering is used.}
\label{fig:smallcomp}
\vspace{-0.8cm}
\end{figure}

Next the effect of the $1/M^2$ current corrections is studied.
Fig.~\ref{fig:smallcomp} shows ratios of the $1/M$ and $1/M^2$ current 
corrections to the zeroth order  current (Eq.~\ref{eq:m0}) at the mass of the $B$ 
meson. The ratio of the $O(1/M)$ operator is $\sim 10\%$, and of the 
$O(1/M^2)$ is $\sim 2\%$. After including renormalization factors in the ratios, the $1/M$ 
correction is still $\sim 10\%$ ~\cite{junkohere}, whereas 
the renormalization  constant for the $1/M^2$ current corrections has not been
calculated yet.

The renormalized matrix elements are obtained from a matching
calculation to full QCD. For NRQCD heavy quarks, two such 1-loop
calculations for operators through $1/M$ exist. In Ref.~\cite{junko}
the full operator mixing at $O(\alpha)$ has been calculated for clover light
quarks.  This calculation shows that there is a significant mixing
with an $O(\alpha \Lambda_{QCD}a)$ current correction.
Ref.~\cite{ishikawa} presents a calculation for Wilson light quarks
which takes only the diagrams with vanishing external momenta into
account.  These give only the contributions of Eq.~(\ref{eq:m0}) and
Eq.~(\ref{eq:m1}) on the renormalization of the zeroth order current
Eq.~(\ref{eq:m0}). The neglected terms are however
significant~\cite{junkohere}.  Also note that for Wilson light quarks
the full mixing calculation at one loop contains an infrared
divergence unless the $O(\alpha \Lambda_{QCD}a)$ correction is
neglected.

\begin{table*}[hbt]
\setlength{\tabcolsep}{0.2pc}
\newlength{\digitwidth} \settowidth{\digitwidth}{\rm 0}
\catcode`?=\active \def?{\kern\digitwidth}
\caption{Summary of results on $f_B$ from moving (non-static) heavy quarks. 
For the results marked with $^\ast$, I have included the renormalization 
constants from Ref.~\protect\cite{junko}. ``t.i. clover'' stands for 
tadpole-improved clover. The results 
from Refs.~\protect\cite{hiroshima,draper,stlouis,saraprogress,JLQCD}
and the systematic errors from~\protect\cite{fermilab} are preliminary.}
\label{tab:summary}
\begin{tabular*}{\textwidth}{@{}|l@{\extracolsep{\fill}}|l|l|c|c|l|}
\hline
\multicolumn{1}{|l|}{group} &
\multicolumn{1}{c|}{heavy quark } &
\multicolumn{1}{c|}{light quark } &
\multicolumn{1}{c|}{$\beta$ } &
\multicolumn{1}{c|}{$a^{-1}$ from} &
\multicolumn{1}{l|}{$f_B$[MeV]} \\
\hline
\multicolumn{6}{|c|}{NRQCD heavy quarks, quenched lattices} \\
\hline
Hiroshima~\protect\cite{yamada} & NRQCD & Wilson & 5.8 & $m_\rho$ 
                                 & $202(17)(^{+11}_{-68})(18)$     \\
          &       &         &   &   & $210(8)(^{+11}_{-71})(6)^\ast$\\
Draper {\em et al.}~\protect\cite{draper} & NRQCD  & Wilson& 6.0  
& $m_\rho$ & $232(11)(^{+8}_{-64})(6)^\ast$ \\
SGO~\protect\cite{sgo_quenched}    & NRQCD & t.i. clover &6.0 & $m_\rho$& 
183(30)(28)(8) \\
GLOK~\protect\cite{stlouis}
    & NRQCD & t.i. clover &6.0 & $m_\rho$ & $149(12)(^{+22}_{-5})(9)$ \\
\hline
\multicolumn{6}{|c|}{NRQCD heavy quarks, $n_f = 2$ lattices from HEMCGC} \\
\hline
SGO~\protect\cite{saraprogress} & NRQCD & t.i. clover & 5.6 & $m_\rho$ &
$156(4)(^{+36}_{-4})(11)$ \\
\hline
\hline
\multicolumn{6}{|c|}{Relativistic heavy quark actions, quenched
lattices} \\
\hline
APE~\protect\cite{APE} & clover ($c_{SW}$=1)& clover ($c_{SW}$=1)
& 6.2 & $f_\pi$ &      180(32) \\
Fermilab~\protect\cite{fermilab} & t.i. clover & t.i. 
clover &5.7, 5.9, 6.1& $f_K$  & 156(35) \\
JLQCD~\protect\cite{JLQCD} & t.i. clover & t.i. clover &5.9, 6.1, 6.3& $m_\rho$ 
& $163(^{+18}_{-20})$ \\
MILC~\protect\cite{MILC}  & Wilson & Wilson &5.7, 5.85, 6.0, 6.3, 6.52& 
$f_\pi$   & $153(^{+40}_{-16}) $   \\
UKQCD~\protect\cite{UKQCD} & clover ($c_{SW}$=1) & clover ($c_{SW}$=1) &  
6.2& $m_\rho, \sqrt{\sigma}$ &  $160(^{+53}_{-20})$ \\
\hline
\end{tabular*}
\vspace{-0.5cm}
\end{table*}
\begin{figure}[t]
\vspace{-5.6cm}
\centerline{\hspace{1.2cm}\epsfxsize=9.5cm
\epsfbox{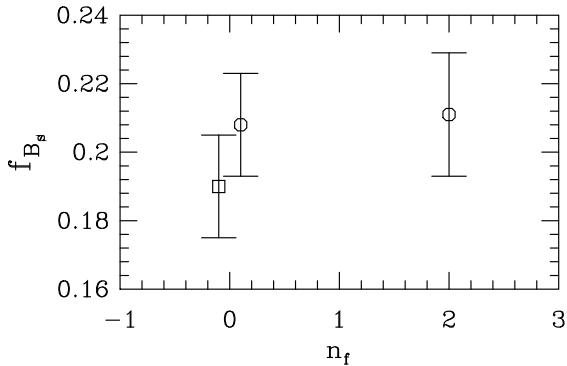}}
\vspace{-2.4cm}
\caption{Comparison of $f_{B_s}$ from NRQCD, using configurations with $n_f
= 0$ and $n_f=2$~\protect\cite{saraprogress}. Quenched results using a $1/M$ 
action~\protect\cite{sgo_quenched} (circle) and a $1/M^2$ 
action~\protect\cite{stlouis} (square)
are shown. The lattice spacing has been fixed from $m_\rho$ ($a^{-1} \simeq
2$ GeV). For better comparison, the $1/M^2$ current corrections are 
not included in the result with the $1/M^2$ action. }
\label{fig:qdyn}
\vspace{-0.3cm}
\end{figure}

In Table~\ref{tab:summary}, a summary of $f_B$ is given;  NRQCD data includes
renormalization constants. I give two results
for the Hiroshima group: (i) from Ref.~\cite{yamada}, using the renormalization
constant determined by them, and (ii) using the full
perturbative $Z$'s (but neglecting any $O(\alpha \Lambda_{QCD} a)$ terms) at
$O(1/M)$~\cite{junko} on their raw data from Ref.~\cite{hiroshima}.
To obtain a renormalized value for $f_B$ from the raw data provided by 
Draper {\em et
al.}~\cite{draper}, I also use the renormalization constant
from~\cite{junko}.  However, note that these $Z$'s are for a slightly
different action than used in Refs.~\cite{hiroshima,draper}.

The first error bar in the NRQCD results in 
Table~\ref{tab:summary} is statistical (including fitting uncertainties where 
applicable). The second error estimates the uncertainty in the
determination of $a^{-1}$. For clover light quarks, the 
upper error bound comes from determining $a^{-1}$ from $f_\pi$; the lower 
bound  from either $\sqrt{\sigma}$~\cite{stlouis}, or
from the statistical error in $m_\rho$~\cite{saraprogress}.  
Ref.~\cite{sgo_quenched} used an $a^{-1}$ of 2 GeV with an error of 0.2 GeV.
For Wilson light quarks, the upper
error bound comes from the error in $a^{-1}$ from
$m_\rho$, the lower error bound from the determination of $a^{-1}$ from
  $\sqrt{\sigma}$.  
Here, the difference between 
$a^{-1}$ from $m_\rho$ and $\sqrt{\sigma}$ at $\beta = 6.0$ is 
$\sim 20\%$, whereas for the clover action they are roughly consistent. 
The third error bar is an estimate of 
higher order contributions in perturbation theory, obtained by using 
$\alpha_P$~\cite{alphapaper}
evaluated  at $q^\ast = 1/a$ and at $q^\ast = \pi/a$ in the $Z$'s, and, for
calculations including only $O(1/M)$ corrections it includes an estimate of 
the $1/M^2$ contributions.

Table~\ref{tab:summary} also lists results on $f_B$ from relativistic
heavy quark actions. For more details on this data see the review by
T.~Onogi~\cite{onogi}.  It is encouraging to note that the result with
NRQCD heavy and clover light quarks~\cite{stlouis} agrees well with
those results with relativistic heavy quarks where extrapolations to
$a\rightarrow 0$ have been done~\cite{fermilab,JLQCD,MILC}. 
The results with NRQCD heavy and Wilson light quarks~\cite{hiroshima,draper} 
in Table~\ref{tab:summary} are high ($> 200$ MeV). This is mainly due to the 
choice of
the lattice scale; using $a^{-1}$ from $\sqrt{\sigma}$ instead of Wilson
$m_\rho$
yields a result close to~\cite{stlouis,fermilab,JLQCD,MILC}. Wilson 
light quarks give rise to considerable discretization effects 
$O(\Lambda_{QCD})$, and if they are used in conjunction with NRQCD heavy  
quarks, these effects cannot be removed by extrapolating to
$a \rightarrow 0$ (which is in principle possible e.g. in~\cite{MILC}). 

Since $f_B$ from NRQCD cannot be extrapolated to $a \rightarrow 0$, it is
important to understand the discretization effects. There exist only 
preliminary results for $f_{B_s}$ as reported by J. Hein {\em et al.}~\cite{hein}. 
They  find slight scaling violations: $f_{B_s}$
decreases by $\sim 2\sigma$ between $\beta = 5.7$ and $6.0$. To make a clear
statement about discretization effects, at least one other calculation at 
smaller $a$ is required. 

Another important question is how quenching affects the heavy-light
decay constant. In Fig.~\ref{fig:qdyn}, $f_{B_s}$ from two
quenched calculations~\cite{sgo_quenched,stlouis} is compared with a result
using $n_f = 2$ lattices~\cite{saraprogress}.
To accommodate the fact that the strange quark masses in the two quenched 
calculations have been determined by either setting $K$ or $K^\ast$ mesons to 
their physical value, the systematic error from various methods to fix the
strange quark mass have been included in the error bars.
Existing data in Fig.~\ref{fig:qdyn} show no significant
deviation between the quenched and the dynamical results.
\subsubsection*{Acknowledgements}
I am grateful to the members of the GLOK Collaboration, in particular 
T.~Bhattacharya, C.~Davies, R.~Gupta, and J.~Shigemitsu, for many helpful 
discussions.

\end{document}